\documentclass[letter]{aa} 
\usepackage{graphicx}
\usepackage{natbib}
\usepackage{txfonts}
\newcommand \hi  {\ion{H}{i}}
\newcommand{\mhz}{\ensuremath{\,\mathrm{MHz}}~}
\newcommand{\kms}{km\,s$^{-1}$}
\newcommand{\khz}{\ensuremath{\,\mathrm{kHz}}~}
\newcommand{\ra}{$\alpha$}
\newcommand{\dec}{$\delta$}
\newcommand{\VLSR}{$\upsilon_{lsr}$}
\newcommand{\unith}{10$^{18}$ cm$^{-2}$}
\newcommand{\FWHM}{$ \Delta \upsilon_{1/2}$}
\newcommand \NH  {$N_{\rm HI}$}
\newcommand \msun{M$_{\sun}$}
\newcommand{\lfd}{\hbox{$^{{\rm d}}$}}
\newcommand{\lfh}{\hbox{$^{{\rm h}}$}}
\newcommand{\lfm}{\hbox{$^{{\rm m}}$}}
\newcommand{\lfs}{\hbox{$\mkern-4mu^{\prime\prime}$}}

\newcommand{\lfarcm}{\hbox{$^\prime$}}
\newcommand{\lfarcs}{\hbox{$^{\rm s}$}}

\begin{document}
 
   \title{An enigmatic  \hi~ cloud.}
   \subtitle{}

   \author{L. Dedes \inst{1} \and C. Dedes \inst{2} \and P.W.M Kalberla \inst{1}}

   \institute{Argelander Institut f\"ur Astronomie (AIfA), University of Bonn,
              Auf dem H\"ugel 71, 53121 Bonn\\
              \email{ldedes@astro.uni-bonn.de, pkalberla@astro.uni-bonn.de} \and
              Max-Planck-Institut f\"ur Radioastronomie, Auf dem H\"ugel 69,
              53121 Bonn }

   \date{Received ------; accepted --------}

 
  \abstract
   {}
   {The  discovery of an \hi~ cloud with peculiar properties at
    equatorial coordinates (J2000) \ra=07\lfh49\lfm, \dec=04\lfd30\lfarcm~ is presented.}
   {The \hi~ object was  detected at 21cm using the 300-m NAIC Arecibo\thanks{The Arecibo Observatory is part of the National Astronomy and Ionosphere Center, which is operated by Cornell University under a cooperative agreement with the National Science Foundation}
     telescope. Subsequent follow-up high-resolution observations with the NRAO
     Very Large Array\thanks{The National Radio Astronomy Observatory is a
       facility of the National Science Foundation operated under cooperative
       agreement by Associated Universities, Inc. } (VLA) telescope at L-Band
     revealed more details about its morphology.}
   {The most intriguing aspect of the object is the clear velocity
         gradient of 1\kms, which is present in the data, an indication 
of either rotation or expansion. The gas appears to be cold, and its
     morphology is somewhat elliptical with clumpy substructure. Assuming
         disk  rotation, the dynamical mass could be determined as a function of distance.}
   { Depending on the exact nature of the velocity gradient in the \hi~
       cloud, we can reach some preliminary conclusions about the nature of the
       object. Expansion would imply association with a circumstellar
   envelope of an evolved AGB star, while in the case of rotation, a comparison between
the visible and the dynamical mass can lead to some preliminary conclusions. A
common feature of those conclusions is the presence of a
gravitational potential well, which is required to account for the rotation of the trapped \hi~
gas. This potential well could be associated with a dark galaxy or some
other exotic object.}  

   \maketitle
%

\section{Introduction}

 During the past decade, the introduction of new state-of-the-art instruments
like the new ALFA multi-pixel receiver of the Arecibo telescope and a superb 
\hi~ all sky survey i.e. the Leiden/Argentinian/Bonn (LAB) survey
\citep{KALBERLA2005AA} has led to a renaissance in Galactic \hi~astronomy. In this 
context, we used the Arecibo 300-m Radio-telescope to survey a field of 17\degr x
5\degr at (l,b)$\sim$(218\degr,15\degr) in the sky for \hi~ halo clouds (Dedes et al; in prep.) with properties similar to
the ones detected by the GBT 100-m telescope \citep{LOCKMAN2002ApJ}. During these
observations, an \hi~ cloud with peculiar properties was detected.

In this short letter we would like to report its properties and discuss its
 nature. The layout is the following. In Sect. \ref{sec_1}, the single
dish Arecibo and the VLA observations are discussed, in Sect. \ref{sec_2}, the
observational properties of the \hi~ cloud  presented, and in
Sect. \ref{sec_3}, possible interpretations regarding the nature of the cloud  discussed.

\section{Observations}
\label{sec_1}

The \hi~ line observations at the Arecibo telescope were done during August 2006.  As front-end, the ALFA multi-beam
receiver was used at a wavelength of 21cm in conjuction with the GALSPECT
spectrometer, which has a bandwidth of 7.14\mhz
and 8192 channels. An area of 17\degr x 5\degr~ was mapped around the
coordinates \ra=08\lfh00\lfm00\lfs, \dec=02\lfd45\lfarcm00\lfarcs. The Arecibo data were reduced using the IDL
routines developed by C. Heiles and J. Peek \citep{STANIMIROVI2006ApJ}. The result was an
image cube with an angular resolution of $\sim$4\arcmin~  and a velocity
resolution of 0.18\kms.\\

The follow-up observations of the cloud took place in July 2008 at the
VLA. The \hi~ cloud was observed for $\sim$ 6 hours in the
L-band with the array in D-configuration. A single IF was used as backend with
a bandwidth of  781\khz and  512 channels. With this configuration we have a  
 channel separation of 1.5\khz and a velocity separation of 0.3\kms.
This is the narrowest possible bandwidth to use for avoiding the
aliasing appearing in the EVLA-EVLA baselines when using a combination between
the 11 VLA and the 16 new  EVLA antennas.\\  

The data were reduced using the AIPS\footnote{http://www.aips.nrao.edu/aips\_faq.html} package, including the
task FXALIAS to correct the aliasing in the EVLA-EVLA baselines. As a flux
calibrator, the source 0137+331 was used, while the close-by source 0739+016
was taken as a phase calibrator. The dirty cube had an rms noise of
$\sim$5mJy/beam. After applying a continuum subtraction, the dirty cube was
de-convolved using the CLEAN Clark algorithm \citep{CLARK1980AA}. 

\begin{figure*}
\centering
\includegraphics[angle=-90,scale=0.323]{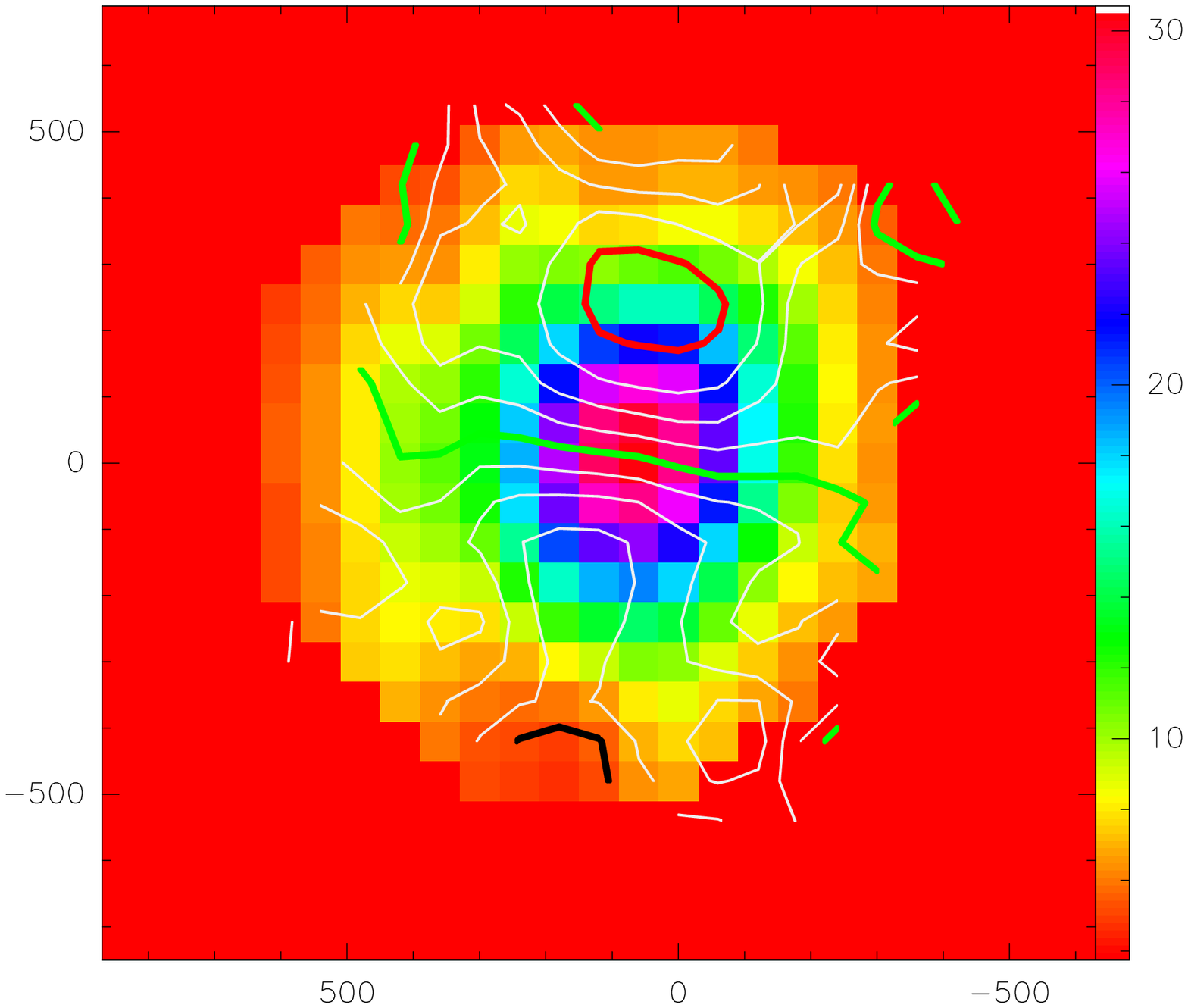}
\includegraphics[angle=-90,scale=0.283]{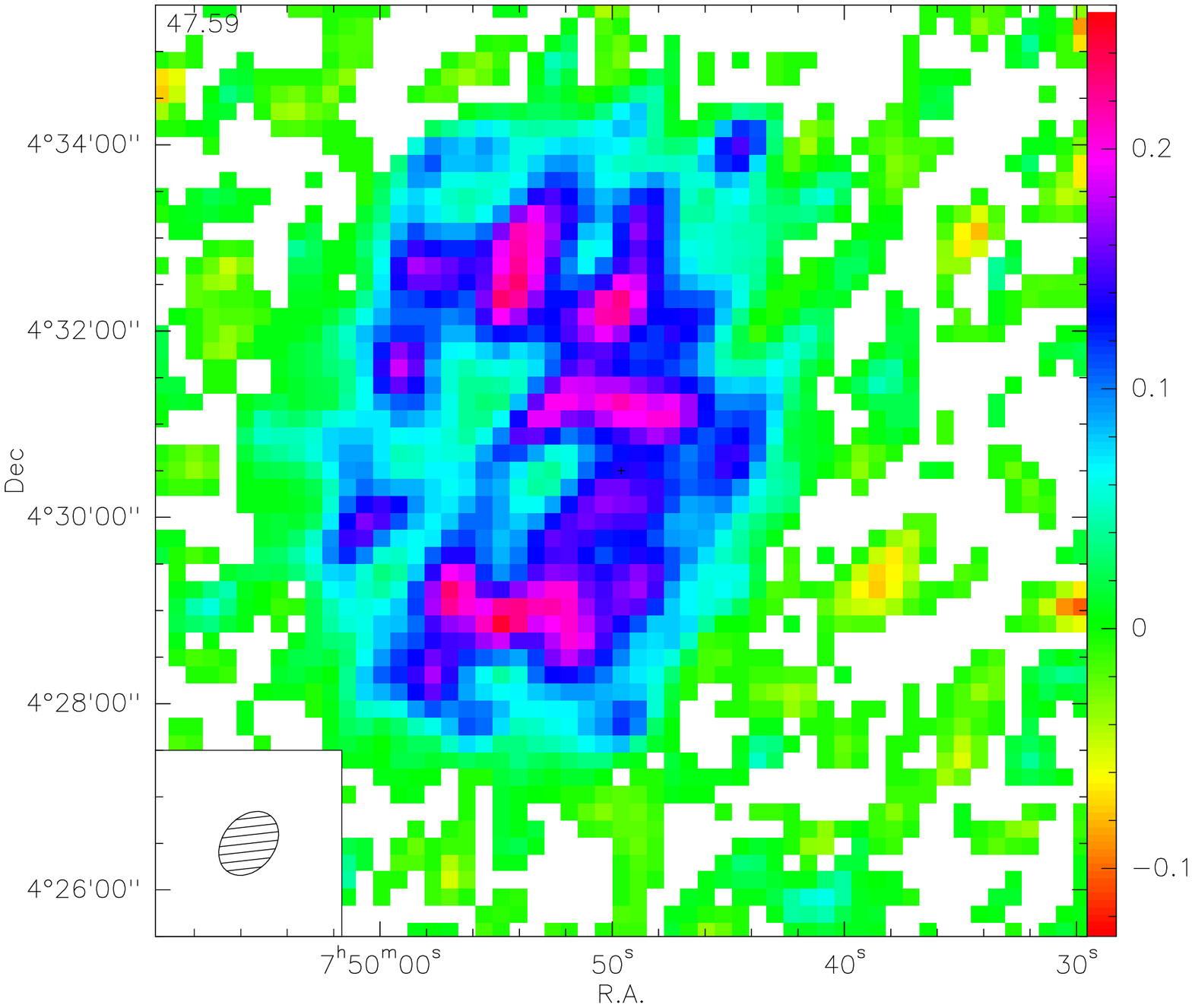}
\includegraphics[angle=-90,scale=0.283]{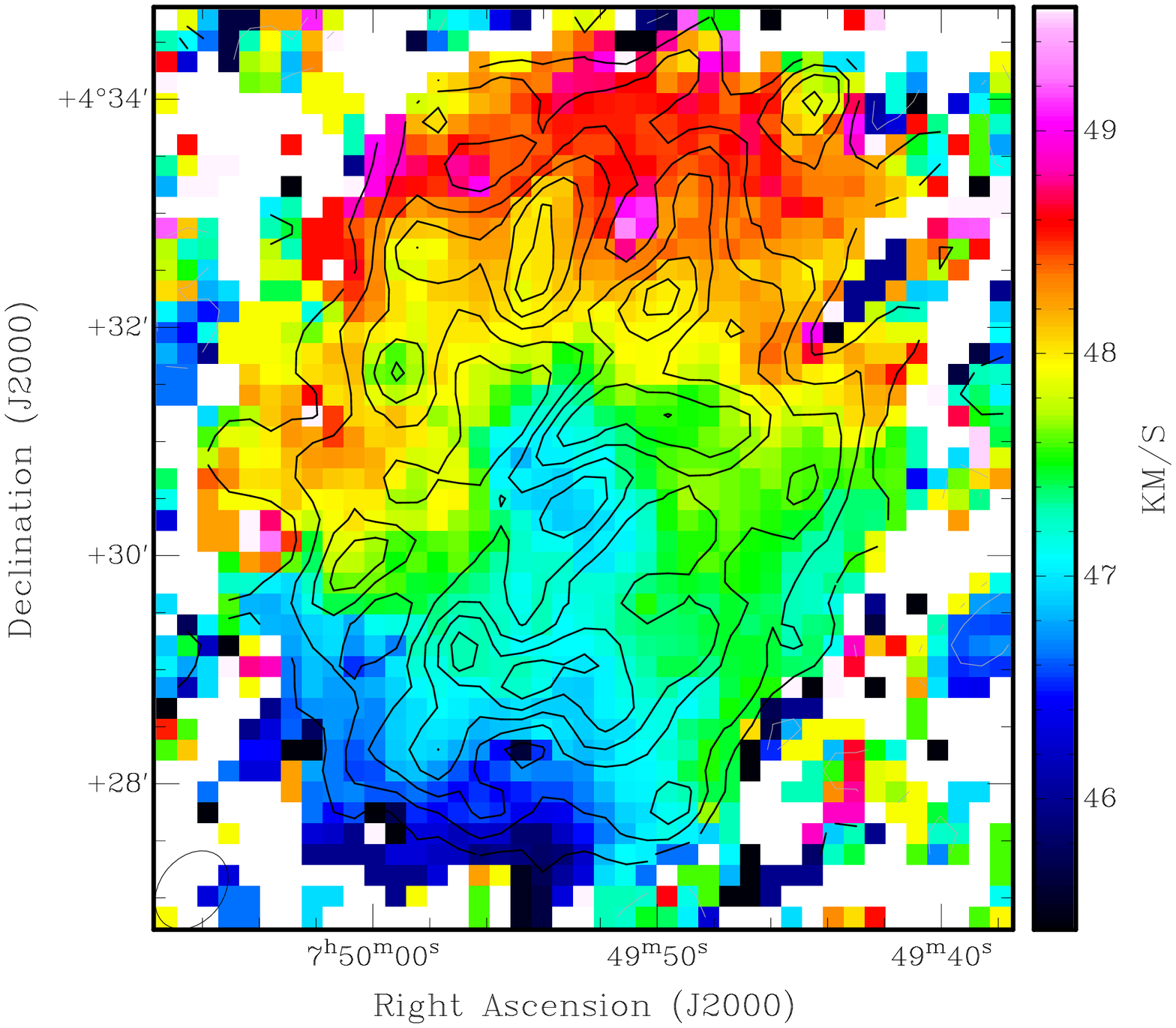}
\caption{Left: a) A 0 moment map of the Arecibo data showing the
      \hi~cloud. Emission outside the object was blanked. The transfer function
      is linear with units of K\kms. As contours, the 1st moment map is
      over-plotted. The contour levels are from \VLSR=47.2\kms~(black) to
      \VLSR=48.00\kms~(red) in steps of 0.1\kms, with 47.6\kms~ marked as a
      green contour. Middle: b) A
  0-moment map of the VLA observation. The transfer function is linear. The units are in Jy/Beam~\kms
. Right: c) A 1st moment map of the VLA observation, depicting velocity from
  45.4\kms~ to 49.6\kms in step of 0.3\kms. The 0 moment  has been overlaid as a contour map with
  contour levels from 10\% up to 90\% of the peak with a step of 10\%.}
\label{0915fig1}
\end{figure*}

The cloud was also observed with  the 295 channel Large Bolometer Camera
(LABOCA) \citep{SIRINGO2007}  at the Atacama Pathfinder Experiment (APEX\footnote{APEX is a collaboration between
the Max-Planck-Institut f\"ur Radioastronomie, the European Southern
Observatory, and the Onsala Space Observatory.}) telescope  during August and
December of 2007. LABOCA operates at 870$\mathrm{\mu}$m with an angular
resolution of 19\arcsec. The total time spent on source was 1.5 hours,
reaching an rms of 5 mJy/beam with an average zenith opacity of 0.2 at 870$\mathrm{\mu}$m. The data were
reduced using the LABOCA reduction package BOA (Schuller et al., in prep). 

\section{Results}
\label{sec_2}
\subsection{Arecibo}

 In the Arecibo data, a spherical \hi~ object with an average
 angular size of 6.4\arcmin at \ra=07\lfh49\lfm49.613\lfs~ and
 \dec=04\lfd30\lfarcm30\lfarcs~ with a \VLSR=47.6\kms~ was detected. It has a column
 density of $N_{peak}=60\times$ \unith~ and average \FWHM~ line width of 3.4$\pm0.18$\kms.
 Figure \ref{0915fig1}a)  shows a zero-moment map at the position of the clump with the 1st moment
 map overlaid as contours. The contours of the velocity field close to the
 peak of the emission show a velocity gradient of 0.8$\pm0.18$\kms~ for the \hi~ object. The parallel lines of the velocity field
 can be interpreted as a signature of a rotating disk or a uniform expanding shell. While
 the spectral resolution is very good, the angular resolution is not good
 enough to provide more information.\\

\subsection{VLA results}
  \begin{figure}
\includegraphics[angle=270,scale=0.7]{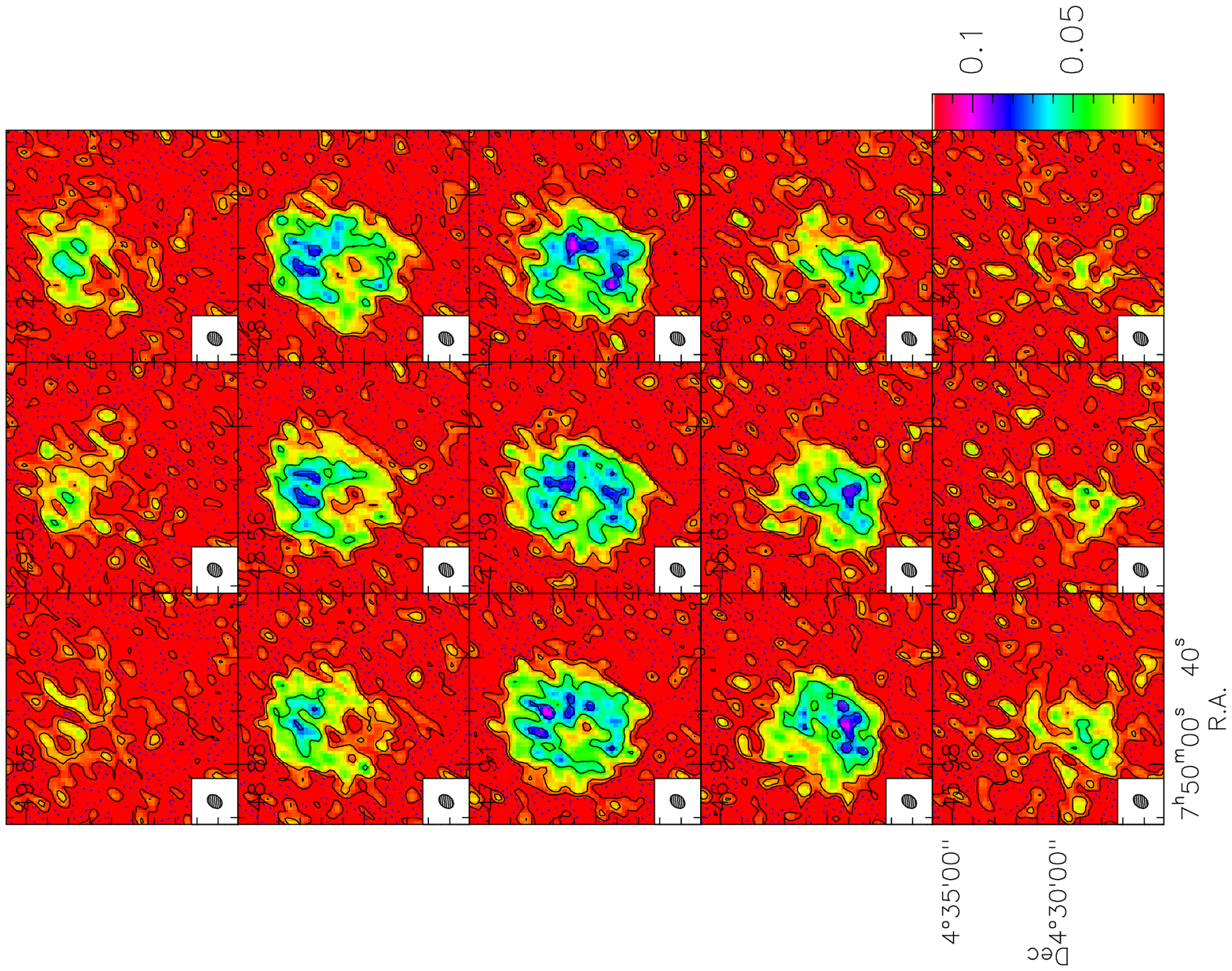}
\caption{Series of channel maps for the VLA data from \VLSR=49.85\kms  to
  \VLSR=48.34\kms.The color scale is r from from 5mJy\kms to 120mJy\kms
  with a linear transfer function.  The contour levels
are plotted from -1$\sigma$ onwards in steps of 1$\sigma$, 3$\sigma$, 9$\sigma$,
15$\sigma$, 21$\sigma$.}
\label{0915fig2}
\end{figure}

The VLA observations confirmed the detection of the velocity gradient of
    the  \hi~ cloud seen with the Arecibo telescope and gave more
information regarding the spatial structure of the object. 

In the consecutive channel maps shown in Fig \ref{0915fig2}, the
  velocity gradient in the cloud is clearly seen. The high-resolution observations revealed a number of interesting features in the
  object:

\begin{figure}
\centering{
\includegraphics[angle=-90,scale=0.3]{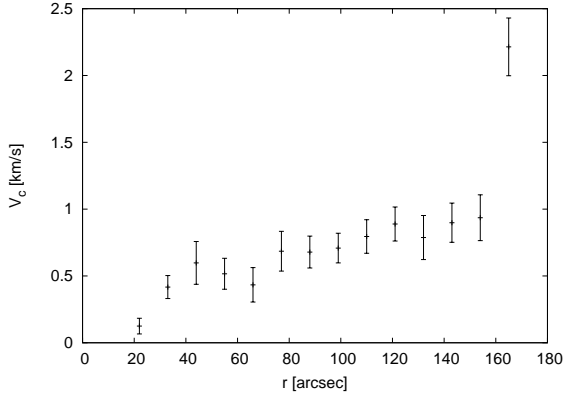}}
\caption{The rotation curve of the \hi~cloud as extracted using the 0-moment
  and the 1st-moment maps. $V_{c}$ is the circular velocity in \kms, r  the
  radius of the clouds from the center, in arcseconds.}
\label{0915fig3}
\end{figure}

\begin{itemize}

\item  In the  0-moment map shown in Fig.\ref{0915fig1} b), the object has an elliptical
  shape in contrast to the spherical shape shown in the Arecibo data. The region where \hi~ emission
  is present can be fitted  with an ellipse of major axis $A=330\pm20$\arcsec,
  minor axis $B=230\pm14$\arcsec, and position angle
  $p.a=-14\pm7$\degr. In the 1st-moment map in Fig.\ref{0915fig1}c, the velocity field of the \hi~cloud is similar to a
  disk-like object rotating as a solid body. The same result is also inferred when the task {\it{Velfit}} of
  MIRIAD\footnote{http://www.atnf.csiro.au/computing/software/miriad/} is used
  to extract a rotation curve from the 0 moment and the 1st-moment maps of the
  \hi~cloud. This is seen in Fig. \ref{0915fig3}. Nevertheless, the
  D-configuration data do not have enough spatial and spectral resolution to
  model  the cloud and exclude expansion sufficiently. The total velocity gradient over the cloud is
  $2\pm0.3$\kms, which would imply an apparent rotational or expansion
      velocity of  $\sim$1\kms.\\

\item  From the 0-moment map in Fig. \ref{0915fig1}b), it can be seen that the \hi~ gas
 does not have a uniform  distribution, which is also seen from the channel
 maps in  Fig. \ref{0915fig2}. The gas has a shell-like filamentary structure, with  several  bright peaks. 
Those have peak column densities from $1\times 10^{20}$cm$^{-2}$ to 1.8$\times 10^{20}$cm$^{-2}$. 
The filaments must be very cold, since they have a line width
\FWHM$\sim2.3$\kms. The \hi~ emission  contained in the peaks
 can be used   to calculate the total visible \hi~ mass. Assuming that the system is at
  a distance $d$ and is optically thin, the total visible
  \hi~ mass as a function of distance, using the average column
  density for each of the peaks, is given by

\begin{equation}
 \label{eq1}   
  \frac{M_{\hi}}{M_{\sun}}=7\cdot10^{-7}\cdot \frac{d^{2}}{pc^{2}}~.
 \end{equation}

 \item In the comparison between the 0-moment and the 1-st moment maps in
   Fig.\ref{0915fig1} c), the last important feature is revealed. At position
 \ra=07\lfh49\lfm53\lfs, \dec=04\lfd30\lfarcm19\lfarcs, there is a column
 density minimum with \NH=3$\times10^{19}$cm$^{-2}$~ at the same position where
 there is a discontinuity in the velocity field. This apparent hole in the
 emission seems to be close to the center of rotation.
\end{itemize}
 
\begin{figure}[h]
\centering{
\includegraphics[angle=-90, scale=0.3]{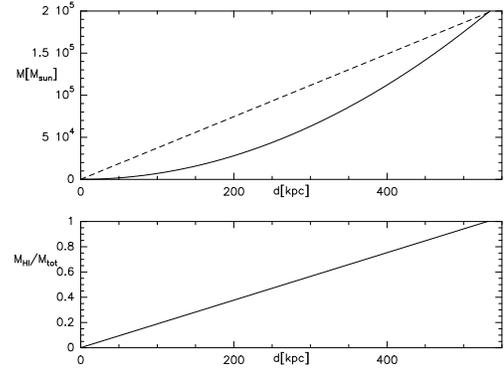}}
\caption{Top : A diagram comparing  the total visible \hi~ mass 
  $M_{\hi}$ (black line) of the cloud measured from  the VLA data and the
  dynamical mass $M_{tot}$ (dashed line) as a function of distance d (kpc). Bottom : Diagram giving the ratio of the
  $M_{\hi}/M_{tot}$ of the cloud as a function of distance d (kpc).}
\label{0915fig4}
\end{figure}

\subsection{Dynamical Mass}

 If the velocity field seen in Fig. \ref{0915fig1} c) can be attributed  to
     a rotating disk following solid-body rotation, the
  dynamical mass can be estimated easily. We assume an apparent rotational velocity
$\sim$1\kms, a disk of inclination $i\sim45$\degr, and an angular radius of
$s=165$\arcsec. At a distance of $d$, the dynamical mass is  

\begin{equation}
\frac{M_{tot}}{M_{\sun}}=0.37\cdot\frac{d}{\mathrm{pc}}~. 
\label{eq2}
\end{equation}

Assuming a uniform rotating disk, the above equation implies an angular
momentum $J=1.3\time10^{46}\times (d/{\rm{pc}})^{2}$m$^{2}\cdot$kg/s. The
distance where the dynamical mass $M_{tot}$ is equal to the visible \hi~
mass is $d_{m}$=530kpc. This implies a total mass of
$M_{tot}$=1.97$\times10^{5}M_{\sun}$ and a diameter of 850pc. In Fig.\ref{0915fig4},
a comparison between the visible \hi~ mass and the dynamical mass is shown. For any distance smaller than  $d_{m}$, the estimated dynamical
mass is significantly greater than the total observed \hi~ mass. As an example,
if the cloud is located at a distance of 50kpc, this would imply that only  10\% of
the total mass is in \hi. In such a case the dynamical mass would be
$M_{tot}$=1.8$\times 10^{4}M_{\sun}$ and its diameter only 80 pc. 


\subsection{Comparison with data at other  wavelengths}

The 0 moment map of the \hi~ VLA observations was also compared with
observations at different wavelengths. The dust continuum observations conducted
with APEX as part of this work show no detection of cold dust above the rms of
5mJy/beam and attempts during the data reduction to recover possible faint extended
emission were unsuccessful. Assuming a dust-to-gas ratio of 1/150 \citep{KLEIN2005ApJS},
$\beta=2$ \citep{PRIDDEY2001MNRAS} and $k_{\nu}(\nu)=0.04 \times (\nu/250
\mathrm{GhZ})^{\beta}$ \citep{KRUEGEL1994AA}, an upper limit for the H$_2$ column
density is $N_{H_2}=9\times 10^{20}$cm$^{-2}$. It has to be taken into account
that the dust parameters are hard to constrain, since the location of the
cloud is not known and average disk values were taken for the
calculation. Assuming e.g. a 10 times lower dust-to-gas ratio to reflect
changing metallicities in the halo \citep{WAKKER2001ApJ}, the column density would change
accordingly.    \\

In addition, utilizing the NED\footnote{This research has made use of the
  NASA/IPAC Extragalactic Database (NED), which is operated by the Jet
  Propulsion Laboratory, California Institute of Technology, under contract
  with the National Aeronautics and Space Administration.},
HEASARC\footnote{http://heasarc.gsfc.nasa.gov/}, SKYVIEW\footnote{http://skyview.gsfc.nasa.gov/} and IRSA\footnote{This research has made use of the NASA/ IPAC Infrared Science Archive, which is operated by the Jet Propulsion Laboratory, California Institute of Technology, under contract with the National Aeronautics and Space Administration. } databases, the
0 moment map was compared with the publicly available surveys at various wavelengths. No obvious correlation was found
except from the 2MASS survey \citep{SKRUTSKIE2006AJ}, where a star was found at coordinates (J2000) \ra=07\lfh49\lfm53.51\lfs,
\dec=04\lfd30\lfarcm23.69\lfarcs . This is exactly the location where the
hole, i.e the drop in the column density of the 0 moment map is found. The apparent K, H, and J magnitudes are 10.225, 10.237, 10.504, respectively. 

\section{Discussion}
\label{sec_3}
From the observational results presented above, four main 
characteristics of the \hi~ cloud can be identified, which, if intrinsic, can give
us an insight regarding the nature of the object: a) the \hi~ cloud
  exhibits a velocity gradient of $\sim2$\kms. While the velocity field is
  compatible with solid-body rotation, based on the VLA data expansion cannot
  be excluded. b) The overall structure as seen in Fig.\ref{0915fig1} c) of
  the \hi~cloud is elliptical. c) The structure as seen in Fig.\ref{0915fig1} b)  is rather filamentary with a hole in the
middle of the 0 moment and the 1st-moment map. d) No correlation with any emission in other
wavelengths was found except in the 2MASS data, where a star was identified
in the hole region. One has to do spectroscopy on the star to rule out
 a projection effect. Based on the above observations and depending on
     the distance, the nature of the \hi~ cloud can be interpreted in several ways:\\

\begin{enumerate}

\item  Placed at a distance of $d_{m}\sim530$kpc, and assuming rotation, it could be a self-gravitating \hi~
disk. However, the filamentary structure of the cloud excludes this
notion as does the difficulty to form such a system and confine the gas
into a disk-like structure. The visible \hi~ mass in each filament ranges 
from $\approx 9\cdot 10^{-8}\times (d/(\mathrm{pc}))^{2}$\msun~ down to $8\cdot 10^{-9}\times
(d/(\mathrm{pc}))^{2}  $\msun, while the virial mass, assuming a dispersion
similar to the line width of 2.3\kms, is $M_{v}$ $\sim 1.29 \times
d/(\mathrm{pc})$. As a result, in order for the filaments to be
self-gravitating, the object would have to be in far greater distance than
530kpc, but in such a distance, their total mass would be greater 
 than the one required for the apparent rotation.

\item For any distance $d<d_{m}$, as seen in
Fig. \ref{0915fig4}, the total visible \hi~ gas cannot account for the observed
rotation. In such a case, the neutral hydrogen plays the role of a tracer of
gravitational potential. The mass responsible for such a gravitational
potential would range from a few hundred solar masses at a distance shorter
than one kpc up to $4\times 10^{4}$\msun~ at a distance of 100kpc. \\
Placed at a distance of about 50kpc, typical for satellites like the Magellanic
Clouds \citep{CIONI2000AA}, the object would have dimensions similar to  dwarf galaxies
(e.g Sgr dwarf, for d=20kpc [Monaco et al. 2004]\nocite{MONACO2004MNRAS} the radius is r$\sim$30pc).
The associated mass in such a case would be lower than a typical dwarf
galaxy mass, which, using the arguments of \citet{JIMENEZ1997MNRAS} might
indicate that the mass is too low to initiate star formation. It has to be
noted that the dynamical mass obtained could be only a lower limit on the true mass,
since it takes into account the rotation of the visible \hi~ gas, therefore underestimating
a possible contribution of dark matter extending past the radius of the
visible mass. It looks as though the object exhibits properties of
the dark galaxy population that is expected to be present in the Milky Way
halo \citep{JIMENEZ1997MNRAS}. These authors also claim that such an object
should be detected in empty fields within deep \hi~ surveys, as was the case
with the data presented here. The object is isolated in the Arecibo data,
blended only with background faint emission of Galactic origin. Another sign 
for the association of the \hi~object with a galaxy is the extreme morphological similarity the object shows
with the atomic gas disk of the Galactic bulge discussed in \citet{FERRIERE2007AA}.\\

\item At a distance greater than 530kpc, the visible mass would exceed the
  dynamical mass. In such a case, the visible mass would not be gravitationally
  bound and one would not expect to observe such a rotational profile. \\

\item Discussion about a more exotic object like a black hole would require studying
the object in tracers such as X-ray emission.

\item  In the above cases, the apparent velocity gradient was explained as
      a result of solid body rotation, but this is not the only feasible
      explanation. Assuming that the \hi~ cloud has a distance of $400$pc
      \citep[i.e average distance of circumstellar envelopes from][]{GERARD2006AJ},
       it would then have a diameter of D=0.53pc, an \hi~mass
      $M_{vis}=0.11$\msun, and an expansion velocity of $\sim$1\kms.
      All three parameters agree well with the ones found in the \hi~ survey of
      circumstellar envelopes around evolved stars \citep{GERARD2006AJ}.  In
      such a case, the \hi~ cloud could form a shell due to mass
      loss from the star detected in the 2MASS survey. Possible detection of
      molecular gas in the \hi~ cloud would further support this option, since
      most of the envelopes in \citet{GERARD2006AJ} are associated with CO emission.

\end{enumerate}

 Nevertheless, it is difficult to reach an irrefutable conclusion regarding the
   nature of the object. More observations are needed, i.e deep synthesis
   21-cm line observations to accurately determine the extent of the
 \hi~ disk and to distinguish between expansion and rotation, optical
 observations to clarify possible association of the star in the 2MASS survey
 with the \hi~ cloud, and observations of CO to determine a possible molecular
 content of the cloud.

\begin{acknowledgements}
  Leonidas Dedes  was supported by the German
      \emph{Deut\-sche For\-schungs\-ge\-mein\-schaft, DFG\/} project
      number KA1265/5-1. L. Dedes would like to thank
      S. Stanimirovic, J. Peek, and K. Douglas for their help with the Arecibo data
      reduction. C. Dedes was supported by the Studienstiftung des Deutschen
      Volkes and is a member of the International Max Planck Research School
      (IMPRS) for Astronomy and Astrophysics. The authors would like to thank Miguel
      Requena-Torres for his useful corrections. The authors would like to
      thank the anonymous referee for the helpful comments.

\end{acknowledgements}

\bibliography{0915}

\begin{thebibliography}{15}
\expandafter\ifx\csname natexlab\endcsname\relax\def\natexlab#1{#1}\fi

\bibitem[{{Cioni} {et~al.}(2000){Cioni}, {van der Marel}, {Loup}, \&
  {Habing}}]{CIONI2000AA}
{Cioni}, M.-R.~L., {van der Marel}, R.~P., {Loup}, C., \& {Habing}, H.~J. 2000,
  \aap, 359, 601

\bibitem[{{Clark}(1980)}]{CLARK1980AA}
{Clark}, B.~G. 1980, \aap, 89, 377

\bibitem[{{Ferri{\`e}re} {et~al.}(2007){Ferri{\`e}re}, {Gillard}, \&
  {Jean}}]{FERRIERE2007AA}
{Ferri{\`e}re}, K., {Gillard}, W., \& {Jean}, P. 2007, \aap, 467, 611

\bibitem[{{G{\'e}rard} \& {Le Bertre}(2006)}]{GERARD2006AJ}
{G{\'e}rard}, E. \& {Le Bertre}, T. 2006, \aj, 132, 2566

\bibitem[{{Jimenez} {et~al.}(1997){Jimenez}, {Heavens}, {Hawkins}, \&
  {Padoan}}]{JIMENEZ1997MNRAS}
{Jimenez}, R., {Heavens}, A.~F., {Hawkins}, M.~R.~S., \& {Padoan}, P. 1997,
  \mnras, 292, L5

\bibitem[{{Kalberla} {et~al.}(2005){Kalberla}, {Burton}, {Hartmann}, {Arnal},
  {Bajaja}, {Morras}, \& {P{\"o}ppel}}]{KALBERLA2005AA}
{Kalberla}, P.~M.~W., {Burton}, W.~B., {Hartmann}, D., {et~al.} 2005, \aap,
  440, 775

\bibitem[{{Klein} {et~al.}(2005){Klein}, {Posselt}, {Schreyer}, {Forbrich}, \&
  {Henning}}]{KLEIN2005ApJS}
{Klein}, R., {Posselt}, B., {Schreyer}, K., {Forbrich}, J., \& {Henning}, T.
  2005, \apjs, 161, 361

\bibitem[{{Kruegel} \& {Siebenmorgen}(1994)}]{KRUEGEL1994AA}
{Kruegel}, E. \& {Siebenmorgen}, R. 1994, \aap, 288, 929

\bibitem[{{Lockman}(2002)}]{LOCKMAN2002ApJ}
{Lockman}, F.~J. 2002, \apjl, 580, L47

\bibitem[{{Monaco} {et~al.}(2004){Monaco}, {Bellazzini}, {Ferraro}, \&
  {Pancino}}]{MONACO2004MNRAS}
{Monaco}, L., {Bellazzini}, M., {Ferraro}, F.~R., \& {Pancino}, E. 2004,
  \mnras, 353, 874

\bibitem[{{Priddey} \& {McMahon}(2001)}]{PRIDDEY2001MNRAS}
{Priddey}, R.~S. \& {McMahon}, R.~G. 2001, \mnras, 324, L17

\bibitem[{{Siringo} {et~al.}(2007){Siringo}, {Weiss}, {Kreysa}, {Schuller},
  {Kovacs}, {Beelen}, {Esch}, {Gem{\"u}nd}, {Jethava}, {Lundershausen},
  {Menten}, {G{\"u}sten}, {Bertoldi}, {De Breuck}, {Nyman}, {Haller}, \&
  {Beeman}}]{SIRINGO2007}
{Siringo}, G., {Weiss}, A., {Kreysa}, E., {et~al.} 2007, The Messenger, 129, 2

\bibitem[{{Skrutskie} {et~al.}(2006){Skrutskie}, {Cutri}, {Stiening},
  {Weinberg}, {Schneider}, {Carpenter}, {Beichman}, {Capps}, {Chester},
  {Elias}, {Huchra}, {Liebert}, {Lonsdale}, {Monet}, {Price}, {Seitzer},
  {Jarrett}, {Kirkpatrick}, {Gizis}, {Howard}, {Evans}, {Fowler}, {Fullmer},
  {Hurt}, {Light}, {Kopan}, {Marsh}, {McCallon}, {Tam}, {Van Dyk}, \&
  {Wheelock}}]{SKRUTSKIE2006AJ}
{Skrutskie}, M.~F., {Cutri}, R.~M., {Stiening}, R., {et~al.} 2006, \aj, 131,
  1163

\bibitem[{{Stanimirovi{\'c}} {et~al.}(2006){Stanimirovi{\'c}}, {Putman},
  {Heiles}, {Peek}, {Goldsmith}, {Koo}, {Kr{\v c}o}, {Lee}, {Mock}, {Muller},
  {Pandian}, {Parsons}, {Tang}, \& {Werthimer}}]{STANIMIROVI2006ApJ}
{Stanimirovi{\'c}}, S., {Putman}, M., {Heiles}, C., {et~al.} 2006, \apj, 653,
  1210

\bibitem[{{Wakker}(2001)}]{WAKKER2001ApJ}
{Wakker}, B.~P. 2001, \apjs, 136, 463

\end{thebibliography}
\bibliographystyle{aa}

\end{document}